\documentclass[%
 reprint,
 amsmath,amssymb,
 aps,
]{revtex4-2}

\usepackage{graphicx}
\usepackage{dcolumn}
\usepackage{bm}

\newcommand{\gaia}{\emph{Gaia}}

\newcommand{\kms}{\,\ensuremath{\mathrm{km\,s}^{-1}}}
\newcommand{\kmskpc}{\,\ensuremath{\mathrm{km\,s}^{-1}\,\mathrm{kpc}^{-1}}}
\newcommand{\masyr}{\,\ensuremath{\mathrm{mas\,yr}^{-1}}}
\newcommand{\muasyr}{\,\ensuremath{\mu\mathrm{as\,yr}^{-1}}}

\newcommand{\ro}{8.275}
\newcommand{\roerr}{0.034}
\newcommand{\vo}{244}
\newcommand{\voerr}{8}
\newcommand{\omegao}{29.4}
\newcommand{\omegaoerr}{1.0}
\newcommand{\dvc}{2}
\newcommand{\dvcerr}{9}
\newcommand{\Ao}{13.8}
\newcommand{\Aoerr}{4.7}
\newcommand{\Bo}{-15.6}
\newcommand{\Boerr}{4.7}
\renewcommand{\ao}{-7.34}
\newcommand{\aoerr}{0.51}
\newcommand{\atilde}{5.05}
\newcommand{\atildeerr}{0.35}
\newcommand{\vsun}{8.0}
\newcommand{\vsunerr}{8.4}

\begin{document}


\title{A purely acceleration-based measurement of the fundamental Galactic parameters}

\author{Jo Bovy}
 \email{bovy@astro.utoronto.ca}
\affiliation{%
 David A. Dunlap Department of Astronomy and Astrophysics, University of Toronto, 50 St. George Street, Toronto, Ontario, M5S 3H4, Canada
}%

\date{\today}

\begin{abstract}
Klioner et al. have used the \gaia\ EDR3 data to directly measure the solar system's acceleration within the Milky Way using the apparent proper motions of quasars. Here I show that this single absolute acceleration measurement in combination with relative accelerations obtained from pulsar orbital decay allows one to determine \emph{all} of the parameters describing the dynamics of our local Galactic environment, including the circular velocity at the Sun $V_0 = \vo \pm \voerr\,\kms$ and its derivative $V'_0 = \dvc \pm \dvcerr\,\mathrm{km\,s}^{-1}\,\mathrm{kpc}^{-1}$, the local angular frequency, the Oort constants, and the Sun's motion with respect to the LSR. This is the first determination of these parameters that only uses the general theory of relativity without the need for additional assumptions.
\end{abstract}

\maketitle

\emph{Introduction---}A good determination of the local gravitational and velocity fields near the Sun in the Milky Way is of fundamental importance to our understanding of the local matter distribution \cite{BlandHawthorn16a}---including its dark matter content \cite{Bovy12a,Strigari13a,Read14a}---for correcting observations for the effects of the Sun's velocity and acceleration, for predicting the expected annual modulation signal in dark-matter detection experiments \cite{Freese13a}, and for many other applications in astro(-particle )physics. However, our knowledge of the gravitational field within \emph{any} galaxy, including our own, has been fundamentally limited by the fact that gravitational theories (i.e., that of the general theory of relativity, or its usually applicable Newtonian limit) only directly connect accelerations to the mass distribution, while we cannot measure the absolute galactic acceleration of \emph{any} star.

This changed recently, because Klioner et al. \cite{Klioner20a} performed a precise and, more importantly, \emph{accurate} determination of the acceleration of the solar system barycenter with astrometric data from \gaia\ EDR3 \cite{GaiaEDR3,Lindegren20a}. This measurement uses the apparent proper-motion pattern of distant quasars induced by the aberration effect that the acceleration gives rise to. This measurement agrees with earlier determinations using Very Long Baseline Interferometry \cite{Titov11a,Titov18a}, but is based on a set of reference sources that is orders of magnitude larger, allowing an analysis that is much freer of systematics. The Milky Way's gravitational field is the largest contributor to the total acceleration \cite{Bachchan16a} and at the current measurement precision, plausible non-axisymmetric contributions to the acceleration are too small to be detected. The largest non-Galactic contribution comes from the Large Magellanic Cloud \cite{Klioner20a}, but it is only at the few percent level and almost perpendicular to the Galactic contribution. The Klioner et al. measurement is therefore a direct measurement of the centripetal acceleration at the Sun, with magnitude $a_0$, which I express as a proper motion $\tilde{a}$
\begin{equation}\label{eq:atilde_obs}
    \tilde{a} = {a_0\over c} = {V_0^2\over c\,R_0} = \atilde \pm \atildeerr\muasyr\,,
\end{equation}
where $V_0$ and $R_0$ are the local circular velocity and the distance to the Galactic center. This acceleration points towards the Galactic center.

\begin{figure*}[t]
\includegraphics[width=\textwidth]{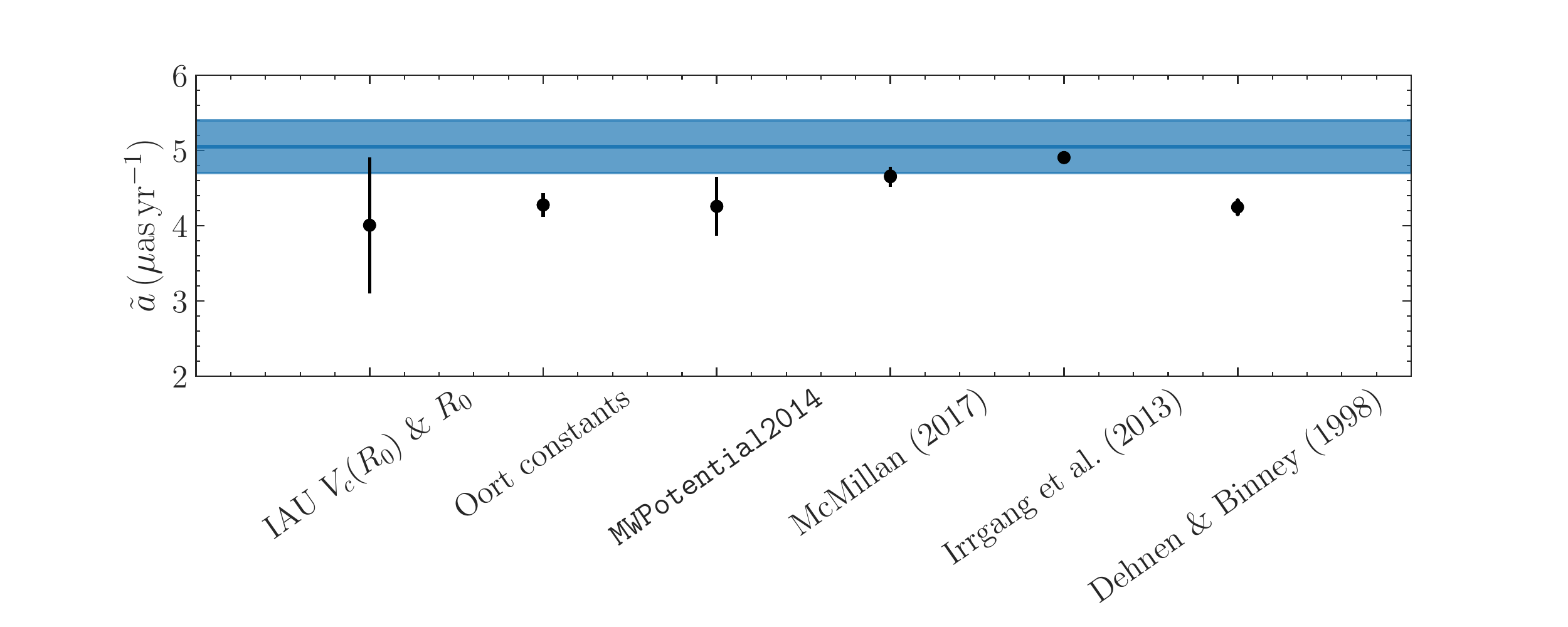}
\caption{\label{fig:atildevslit} Comparison of the \gaia\ EDR3 measurement of the local acceleration, expressed as a proper motion $\tilde{a}$ (see Eqn. \ref{eq:atilde_obs}), with previous indirect determinations and popular Milky Way models.}
\end{figure*}

Galactic gravitational fields are often expressed in terms of the velocity of a circular orbit and the circular velocity at the Sun, $V_0$, is one of the most fundamental parameters of the Milky Way. It is clear that the acceleration measurement in Equation \eqref{eq:atilde_obs} provides a stringent constraint on $V_0$, but to actually obtain $V_0$ (and all of the other fundamental parameters that I will discuss below), we need to combine Equation \eqref{eq:atilde_obs} with a measurement of the Sun's distance to the Galactic center, $R_0$. Luckily, we are fortunate enough to now have a measurement of $R_0$ that essentially only employs fundamental physics, because a highly precise measurement of $R_0$ was obtained from post-Newtonian modeling of the orbit of the star S2 around the Milky Way's supermassive black hole. Precise measurements of S2's orbit using the GRAVITY instrument on the VLT have allowed its gravitational redshift \cite{Gravity18a} and Schwarzschild precession \cite{Gravity20a} to be measured, allowing a precise determination of $R_0$ \cite{Gravity19a}, because $R_0$ connects the observed motion on the sky with the observed line-of-sight velocity projection of the orbit. I use the value for $R_0$ from the recent re-analysis of the GRAVITY data \cite{Gravity21a}
\begin{equation}\label{eq:ro}
    R_0 = \ro \pm \roerr\,\mathrm{kpc}\,,
\end{equation}
but note that the GRAVITY collaboration's preferred value for $R_0$ changes by more than its uncertainties between different analyses (a situation addressed in Ref. \cite{Gravity21a}) and disagrees with the analysis of the gravitational-redshift of S2 by Ref. \cite{Do19a}, who find $R_0 = 7.95\pm0.06\,\mathrm{kpc}$. Together, the measurements in Equations \eqref{eq:atilde_obs} and \eqref{eq:ro} allow the local circular velocity to be determined
\begin{equation}\label{eq:vo}
    V_0 = \left(\vo \pm \voerr\,\kms\right)\,\sqrt{{R_0\over \ro\,\mathrm{kpc}}}.
\end{equation}
For the value of $R_0$ from Ref. \cite{Do19a}, the central value shifts only to $V_0 = 239\kms$.

Another parameter that immediately follows is the angular frequency of the circular orbit at the Sun, $\Omega_0 = V_0 / R_0$, which is constrained to be
\begin{equation}\label{eq:omegao}
    \Omega_0 = \left(\omegao\pm \omegaoerr\kmskpc\right)\,\sqrt{{ \ro\,\mathrm{kpc}\over R_0}}\,.
\end{equation}

However, we can do much more with these measurements if we combine them with a few other key measurements of velocities and relative accelerations in the Milky Way and in fact determine essentially all fundamental parameters describing the local gravitational and velocity field. I explore this in this brief paper. While the resulting measurements do not currently have high precision, they stand out as being the only ones that do not assume dynamical equilibrium or any other method for connecting observed velocities to the gravitational field.\\

\emph{The Sun's motion with respect to the LSR---}Generally a nuisance parameter, the Sun's peculiar velocity with respect to the circular orbit through its location (the LSR) is an important parameter when studying the local velocity distribution and its gradients and is, for better or worse, often used as part of the correction of observed velocities to the Galactocentric reference frame (this is, in fact, unnecessary; however, it is often done). While in the radial and vertical directions, the assumption of axisymmetry allows the Sun's motion to be determined simply as the opposite of the mean velocity of local stars in those directions---because on average these velocities should be zero in the Galactocentric frame---the asymmetric drift makes the determination of the Sun's peculiar velocity in the direction of Galactic rotation, $V_\odot$, much trickier \cite{BinneyTremaine}. The generally accepted value derived from observations of local stars is $V_\odot \approx 12\kms$ faster than Galactic rotation \cite{Schoenrich10a}. However, stellar velocities on larger scales are most easily understood if $V_\odot \approx 25\kms$ \cite{Bovy12a,Bovy15a}. This discrepancy, if real, is an indication that the solar neighborhood as a whole may be perturbed away from a circular orbit by non-axisymmetric forces, at the level of $\approx 10\kms$.

By combining the measurements of $R_0$ and $V_0$ in the introduction with the observed proper motion $\mu_{\mathrm{Sgr\ A}^*}$ of the supermassive black hole at the center of the Milky Way, Sgr A$^*$---assumed to be at rest at the Galactic center---we can now directly determine the Sun's peculiar velocity, because
\begin{equation}
    \mu_{\mathrm{Sgr\ A}^*} = \Omega_0 + {V_\odot\over R_0}\,.
\end{equation}
Combining the latest measurement of $\mu_{\mathrm{Sgr\ A}^*} = -6.411\pm 0.008\masyr$ \citep{Reid20a} with the measurements from the introduction, we get
\begin{equation}
    V_\odot = \vsun \pm \vsunerr\kms\,.
\end{equation}

This measurement is at the lower end of the previously considered range and it agrees with the accepted $V_\odot= 12\pm 2\kms$ from Ref. \cite{Schoenrich10a}. It is about $2\sigma$ off from the global value of $V_\odot \approx 25\kms$ from Ref. \cite{Bovy12a,Bovy15a}. The uncertainty in $V_\odot$ is largely driven by the uncertainty in $\tilde{a}$ and it should therefore improve significantly with future \gaia\ data releases, which will allow us to get to the root of the discrepancy between local and global values.\\

\begin{figure*}[t]
\includegraphics[width=0.7\textwidth]{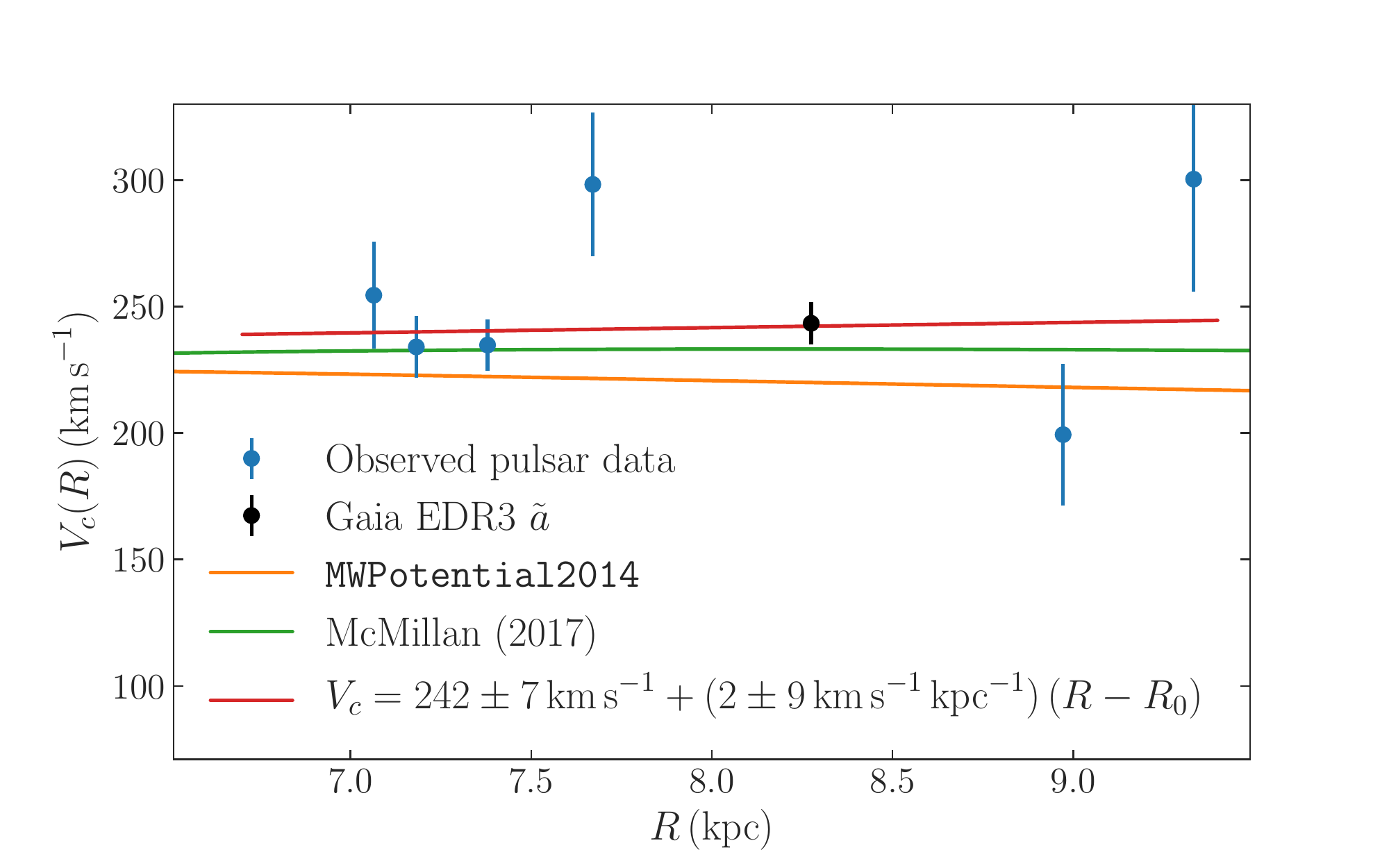}
\caption{\label{fig:rotcurve} Local circular velocity curve determined solely from observed absolute accelerations for the solar system (\gaia\ EDR3; black point) and 6 pulsars relatively close to the Galactic mid-plane whose relative acceleration obtained from pulsar timing can be turned into an absolute determination using the solar system's acceleration. The rotation curve is approximately flat over the 2 kpc surrounding the solar radius. Also shown is a linear fit (red curve) and the predicted rotation curve for two popular models for the Milky Way's gravitational field.}
\end{figure*}

\emph{The local acceleration compared to previous constraints and models---}Before continuing the determination of the local fundamental dynamical parameters, we briefly discuss how the observed acceleration compares to that derived from earlier determinations of $V_0$ and $R_0$ and also how often-used models for the Milky Way's gravitational potential predicted the observed acceleration. This is shown in Fig. \ref{fig:atildevslit}. I compare there to the parameters still recommended by the IAU \cite{Kerr86a}, which have $V_0 = 220\pm 20\kms$ and $R_0 = 8.5\pm 1.0\,\mathrm{kpc}$, but also to the value implied by the measurement of the local Oort constants using \gaia\ DR1 \cite{Bovy17a} after combining it with the measurement of $R_0$ from Equation \eqref{eq:ro}; these both imply much lower values of $\tilde{a} \approx 4$ to $4.3\muasyr$ compared to the observed value, indicating for the Oort constants that they are likely affected by non-axisymmetric streaming motions. There have been, of course, dozens, if not more, determinations of $V_0$ in the recent past that could be cast as measurements of $\tilde{a}$ and I do not intend to review all of them here. Suffice it to say that the observed value of $\tilde{a}$ is at the high end of that implied by previous measurements of $V_0$.

We also compare to some standard models of the Milky Way's gravitational field: the \texttt{MWPotential2014} model from Ref. \cite{galpy} (assuming an uncertainty of $10\kms$ on $V_0$ and $0.1\,\mathrm{kpc}$ in $R_0$), the model from McMillan (2017) \cite{McMillan17}, model I from Irrgang et al. (2013) \cite{Irrgang13a} (which is similar to the often used model of Allen \& Santillan \cite{Allen91a}), and the range predicted by models I though IV of Dehnen \& Binney (1998) \cite{Dehnen98a}. All of these models, except for that of Irrgang et al., predict values for $\tilde{a}$ that are smaller than the observed value, although within the current level of uncertainty, the tension is not very large. We can conclude that for the current measurement, Galactic models are at minor variance with the directly determined acceleration and that future improvement in the direct determination will be a valuable constraint on global Milky Way models.

\emph{Using pulsar orbital decay to directly measure the radial dependence of the Galactic acceleration---}So far I have used only the single, new local acceleration measurement from \gaia\ EDR3. However, while no absolute measurements of the acceleration of any star existed before, measurements of the \emph{relative} acceleration of a small number of stars with respect to the solar system barycenter did exist. These derive from the fact that the observed period changes of binary pulsars have a contribution from the relative acceleration of the pulsar and observer reference frames (in addition to the, often considered more interesting, gravitational-wave contribution) \cite{Damour91a,Weisberg16a}. As recently emphasized by Chakrabarti et al. \cite{Sukanya20a} and Philips et al. \cite{Phillips21a}, assuming that the general theory of relativity holds and, thus, that its contribution to the observed period changes can be determined (albeit with uncertainties), the observed period changes can be turned into measurements of the relative Galactic acceleration at the binary pulsar's location and the Sun. Chakrabarti et al.\ used such relative accelerations derived from observations of 14 binary pulsars to constrain the local gravitational field, while Philips et al. use both spin and orbital periods to do the same. Relative accelerations are a useful constraint on their own, but \gaia's measurement of the local acceleration itself allows these relative accelerations to be turned into a set of absolute accelerations covering a range of distances from the Sun. Complementary to the analysis of Chakrabarti et al., who focus largely on constraining the vertical gravitational field because they are limited to relative accelerations (and the Sun's vertical acceleration is approximately zero), I use the absolute accelerations here to determine the radial behavior of the acceleration or, if you will, the local circular velocity curve.

Chakrabarti et al.\ \cite{Sukanya20a} compiled observed periods and period changes, parallaxes and proper motions (necessary for a purely kinematic contribution to the observed period change \cite{Shklovskii70a}), and period changes expected from the general theory of relativity, for a sample of field binary pulsars. I use their compilation, but remove four pulsars with such large measurement uncertainties that they do not meaningfully constrain Galactic radial accelerations (J0737-3039A/B, J1012+5307, J1603-7202, and J2129-5721) and I add the famous Hulse-Taylor pulsar, B1913+16, which is the only pulsar useful for this exercise at multiple-kpc distances from the Sun. Its parameters are obtained from Ref. \cite{Weisberg16a}, but I use the geometric parallax and proper motion from Ref. \cite{Deller18a}. However, we will find that for the current value and uncertainty of the parallax, the Hulse-Taylor pulsar is in strong disagreement with the expected Galactic acceleration (to a degree that cannot be explained any plausible model), so it will not be useful for constraining the local dynamical parameters. But it's fun and illustrative to include it.

The observed period change $\dot{P}_b/P_b$ (where $P_b$ is the orbital period) is given by \cite{Damour91a}
\begin{equation}
    \left({\dot{P}_b \over P_b}\right)_{\mathrm{obs}} = {1 \over c}\,\vec{x}\cdot(\vec{a}_1-\vec{a}_0) + {\mu^2\,D \over c} +\left({\dot{P}_b \over P_b}\right)_{\mathrm{GR}}\,,
\end{equation}
where $D$ is the distance to the binary pulsar, $\mu$ is the magnitude of its proper motion, $\vec{x}$ is the unit vector pointing to the pulsar from the solar system barycenter, and $\vec{a}_1$ and $\vec{a}_0$ are the Galactic acceleration vectors at the location of the binary pulsar and the Sun, respectively.

We can then compute the line-of-sight acceleration $\vec{x}\cdot\vec{a}_1$ from the pulsar observations (I additionally have to assume a value for the Sun's offset from the Galactic mid-plane, for which I use $Z_\odot =20.8\,\mathrm{pc}$ from Ref. \cite{Bennett19a}), GR effects, and using the measurement of $\vec{a}_0/c$ from Equation \eqref{eq:atilde_obs} (note that I assume axisymmetry, see above). Comparing the computed line-of-sight accelerations for the 11 pulsars to the predicted line-of-sight accelerations in two popular models for the Milky Way's gravitational field, the \texttt{MWPotential2014} model from Ref. \cite{galpy} and the model McMillan (2017; \cite{McMillan17}) reveals excellent agreement. The sole exception is the Hulse-Taylor pulsar (B1913+16), which deviates significantly from the observed value. However, the observed line-of-sight acceleration is strongly influenced by the assumed distance to the pulsar and this distance is highly uncertain. I have used the recent distance derived from a trigonometric parallax \cite{Deller18a}, which placed the pulsar at a surprisingly small distance, and the discrepancy between the Galactic acceleration derived using this distance (taking into account its uncertainty) and the expected value shows that the distance is likely much larger. Note that there isn't any plausible gravitational field that would make the determined acceleration make sense at its assumed distance, because its sign is wrong. Determining a more precise distance to the Hulse-Taylor pulsar would be useful, because as the binary pulsar at the largest distance, it would provide a strong constraint on the Galactic potential.

To determine the local rotation curve, I focus on those pulsars for which the contribution from the vertical acceleration to the line-of-sight projection is negligible compared to that from the radial acceleration. Specifically, I select those quasars for which the relevant term in the projection involving the vertical acceleration is at least five times smaller than the relevant terms involving the radial acceleration, assuming the \texttt{MWPotential2014} model (however, the cut does not significantly depend on what potential model is assumed). This selects 6 pulsars for which I can then determine the radial acceleration and its implied circular velocity at the radius of the pulsar. To account for the slight decrease of the circular velocity with height for these pulsars that are a few hundred parsec from the plane, I correct the circular velocities using Eqn. (15) from Ref. \cite{Bovy12a} using a scale length of 2.5 kpc (note, however, that this correction amounts to only a few $\mathrm{km\,s}^{-1}$). The resulting values are shown in Fig. \ref{fig:rotcurve}. It is clear that while the uncertainties are large, the rotation curve is consistent with being flat over the 2 kpc range surrounding the solar circle. 

We can determine the local slope of the rotation curve by fitting a simple linear model to these data, which is shown by the red line in Fig. \ref{fig:rotcurve}. This gives a local slope of
\begin{equation}
    V'_0 \equiv {\mathrm{d} V_c \over \mathrm{d} R} = \dvc \pm \dvcerr\kmskpc\,,
\end{equation}
which is consistent with being flat. The dependence of this measurement on the assumed value of $R_0$ from Equation \eqref{eq:ro} is very small; even changing $R_0$ to 7.7 kpc does not change it. With this measurement, we can complete our determination of the fundamental Galactic parameters describing the local gravitational field. Traditionally, the local velocity field has been described using the so-called Oort constants $A$ and $B$ \cite{Oort27a,Bovy17a}, which describe the local shear and vorticity of the velocity field and which are directly related to the local angular frequency $\Omega_0$ and the slope of the rotation curve $V'_0$. The implied measurements for the Oort constants are shown together with all of the parameters that I discussed in Table \ref{tab:results}.\\

\begin{table}[t]
\caption{\label{tab:results}%
Values of the fundamental Galactic parameters derived from geometric or fundamental-physics measurements of distances, velocities, and accelerations.}
\begin{ruledtabular}
\begin{tabular}{lcr}
\textrm{Parameter (unit)}&
\textrm{Measurement}&
\textrm{Reference}\\
\colrule
$a_0$ (km\,s$^{-1}$\,Myr$^{-1}$) & $\ao  \pm \aoerr$  & Klioner et al. (2020)\\
$R_0$ (kpc) & $\ro  \pm \roerr$  & Gravity collab. et al.\\
$V_0$ (km\,s$^{-1}$) & $\vo  \pm \voerr$  & Bovy (2020; this paper)\\
$\Omega_0$ (km\,s$^{-1}$\,kpc$^{-1}$) & $\omegao  \pm \omegaoerr$  & "\\
$V'_0$ (km\,s$^{-1}$\,kpc$^{-1}$) & $\dvc  \pm \dvcerr$  & "\\
$A$ (km\,s$^{-1}$\,kpc$^{-1}$) & $\Ao  \pm \Aoerr$  & "\\
$B$ (km\,s$^{-1}$\,kpc$^{-1}$) & $\Bo  \pm \Boerr$  & "\\
$V_\odot$ (km\,s$^{-1}$) & $\vsun  \pm \vsunerr$  & "\\
\end{tabular}
\end{ruledtabular}
\end{table}

\emph{Conclusion---}\gaia\ EDR3's direct measurement of the acceleration of the solar system within the Milky Way galaxy from the apparent proper motion pattern of distant quasars that it induces, is a revolutionary moment for Galactic astrophysics. As I demonstrated here, this single measurement allows us to directly infer most of the fundamental parameters describing the local gravitational and velocity fields using only geometric measurements or measurements that only assume fundamental physics (essentially, the general theory of relativity), without the usual need for additional astrophysical assumptions. These determinations are summarized in Table \ref{tab:results}. While the current precision in these measurements is not that high, this should improve with future measurements of the local acceleration using \gaia\ data, which should get better by a factor of $\approx 3$ in \gaia's fourth data release and $\approx 8$ in its fifth \cite{Klioner20a}. Given that the analysis of \gaia\ DR2 paper has conclusively demonstrated that the local velocity field is strongly affected by non-axisymmetry in the Galactic potential \cite{Antoja18a,Kawata18a,Hunt19a,BlandHawthorn19a,Bennett19a} and thus that any assumption-based inference of the local dark and visible matter distribution based on it is suspect \cite{Banik17a}, such assumption-free, fundamental-physics measurements of the local gravitational field will be of great importance in furthering our understanding of the local matter distribution and its perturbations.\\[10pt]

\begin{acknowledgments}
JB acknowledges financial support from NSERC (funding reference number RGPIN-2020-04712) and an Ontario Early Researcher Award (ER16-12-061). This work has made use of data from the European Space Agency (ESA) mission
{\it Gaia} (\url{https://www.cosmos.esa.int/gaia}), processed by the {\it Gaia}
Data Processing and Analysis Consortium (DPAC,
\url{https://www.cosmos.esa.int/web/gaia/dpac/consortium}). Funding for the DPAC
has been provided by national institutions, in particular the institutions
participating in the {\it Gaia} Multilateral Agreement.
\end{acknowledgments}

\newcommand{\mnras}{Mon.\ Not.\ Roy.\ Astron.\ Soc.}
\newcommand{\aj}{Astron.\ J.}
\newcommand{\apjl}{Astrophys.\ J.\ Lett.}
\newcommand{\apjs}{Astrophys.\ J.\ Suppl.}
\newcommand{\aap}{Astron.\ Astrophys.}
\newcommand{\physrep}{Phys. Rep.}
\newcommand{\araa}{Ann. Rev. Astron. Astrophys.}
\newcommand{\sovast}{Soviet Astronomy}
\bibliography{apssamp}

\end{document}